\documentclass[conference]{IEEEtran}
\IEEEoverridecommandlockouts
\usepackage{cite}
\usepackage{amsmath,amssymb,amsfonts}
\usepackage{algorithmic}
\usepackage{graphicx}
\usepackage{textcomp}
\usepackage{xcolor}

\usepackage{bm}
\usepackage{subfigure}
\usepackage{colortbl}
\usepackage{algorithm}

\newtheorem{theorem}{{Theorem}}
\newenvironment{proof}{{\it Proof:}} {\hfill $\blacksquare$\par} 

\def\BibTeX{{\rm B\kern-.05em{\sc i\kern-.025em b}\kern-.08em
    T\kern-.1667em\lower.7ex\hbox{E}\kern-.125emX}}
\begin{document}

\title{Rate Maximization for RIS-Assisted OAM Multiuser Wireless Communications
\thanks{This work is supported in part by the National Natural Science Foundation of China under Grant 62201427 and the Natural Science Basic Research Program of Shaanxi under Grant 2024JC-YBQN-0642. \textit{(Corresponding author: Liping Liang.)}}
}

\author{\IEEEauthorblockN{Jun Lan$^{\dagger}$, Liping Liang$^{\dagger}$, Wenchi Cheng$^{\dagger}$, and Wei Zhang$^{\ddagger}$}~\\[0.2cm]
\vspace{-10pt}

\IEEEauthorblockA{$^{\dagger}$State Key Laboratory of Integrated
Services Networks, Xidian University, Xi'an, China\\
$^{\ddagger}$School of Electrical Engineering and Telecommunications, The University of New South Wales, Sydney, Australia\\
E-mail: \{\emph{junl@stu.xidian.edu.cn}, \emph{liangliping@xidian.edu.cn}, \emph{wccheng@xidian.edu.cn}, \emph{w.zhang@unsw.edu.au}\}
}

\vspace{-20pt}
}

\maketitle

\begin{abstract}
Conventional multiple-input multiple-out (MIMO) technologies have encountered bottlenecks of significantly increasing spectrum efficiencies of wireless communications due to the low degrees of freedom in practical line-of-sight scenarios and severe path loss of high frequency  carriers. Orbital angular momentum (OAM) has shown the potential for high spectrum efficiencies in radio frequency domains. To investigate the advantage of OAM in multiuser communications, in this paper we propose the reconfigurable intelligence surface (RIS) assisted OAM multiuser (MU) wireless communication schemes, where RIS is deployed to establish the direct links blocked by obstacles between the OAM transmitter and users, to significantly increase the achievable sum rate (ASR) of MU systems. 
To maximize the ASR, we develop the alternative optimization  algorithm to jointly optimize the transmit power and phase shifts of RIS.
The numerical outcomes demonstrate the superiority of our proposed scheme compared to existing methods in terms of ASR.
\end{abstract}
\section{Introduction}
\IEEEPARstart{T}{he} exponential growth in demand for mobile multimedia services has imposed higher requirements on existing communication systems 
\cite{9003618},\cite{1}.
To meet the demand, academics and industry have explored several techniques, such as massive multiple-input multiple-output (MIMO) and integrated sensing and communication. 
Whereas existing conventional MIMO technologies have confronted bottlenecks of significantly increasing spectrum efficiencies of wireless communications due to the low degrees of freedom in practical line-of-sight (LoS) scenarios and severe path loss of high frequency  carriers.

Several academics have garnered growing interest in leveraging the physical characteristics of electromagnetic waves, such as orbital angular momentum (OAM) \cite{2}--\cite{3} to enhance the spectrum efficiency (SE) of wireless communications due to the inherent orthogonality among OAM modes.  
Hence, OAM can be jointly used with conventional orthogonal frequency division multiplexing (OFDM) for high SE of wireless communications \cite{liang2020joint}. 
In \cite{9442905},  an overall scheme was proposed for the multiuser (MU) OAM, achieving significantly higher SE and energy efficiency  compared to  MU-MIMO systems.
Existing research on OAM mainly focuses on LoS scenarios due to the divergence of high-order OAM beams. 

However, the LoS links between the transceivers are vulnerable to being blocked by obstacles in many practical scenarios, thus severely degrading the performance of wireless communications. 
Focused on this issue, reconfigurable intelligence surface (RIS), which has the capability to reconfigure wireless environment and  improve performance of wireless communication systems, has attracted extensive attention  \cite{yang2021reconfigurable}.
RIS can adjust the amplitude and phase of received signals to enhance the strength of useful signals while weakening interference \cite{10330152}.
The RIS aided OAM scheme was investigated to maximize SE while addressing the divergence challenge of OAM beams in point-to-point wireless communications and also enabling obstacle circumvention for maintaining communication capabilities.

Inspired by the advantages of OAM and RIS, we propose the RIS-assisted OAM MU (RIS-OAM-MU) communication scheme, where RIS is deployed to connect the OAM transmitter with multiple OAM users blocked by obstacles, to dramatically enhance the achievable sum rate (ASR) of wireless communications.  To maximize the ASR of our proposed scheme, we develop the alternative optimization algorithm with low computational complexity to joint optimize the transmit power and phase shifts of RIS subject to the constraints of total transmit power and the modulus of phase shifts. The numerical results indicate that RIS-OAM-MU scheme can significantly enhance ASR of scenarios where the transmission is blocked.

The remainder of this paper is organized as follows. The RIS-OAM-MU system  model based on uniform circular array (UCA) is given in Section \uppercase\expandafter{\romannumeral2}. Section \uppercase\expandafter{\romannumeral3} formulates ASR optimization problem and develops the alternative optimization algorithm to maximize the ASR. Numerical results are shown in Section \uppercase\expandafter{\romannumeral4} and conclusions are summarized in Section \uppercase\expandafter{\romannumeral5}.

\section{System Model}\label{sec:sys}
\begin{figure}[htbp] 
\vspace{-0.2cm}
\centering
\includegraphics[width=0.44\textwidth]{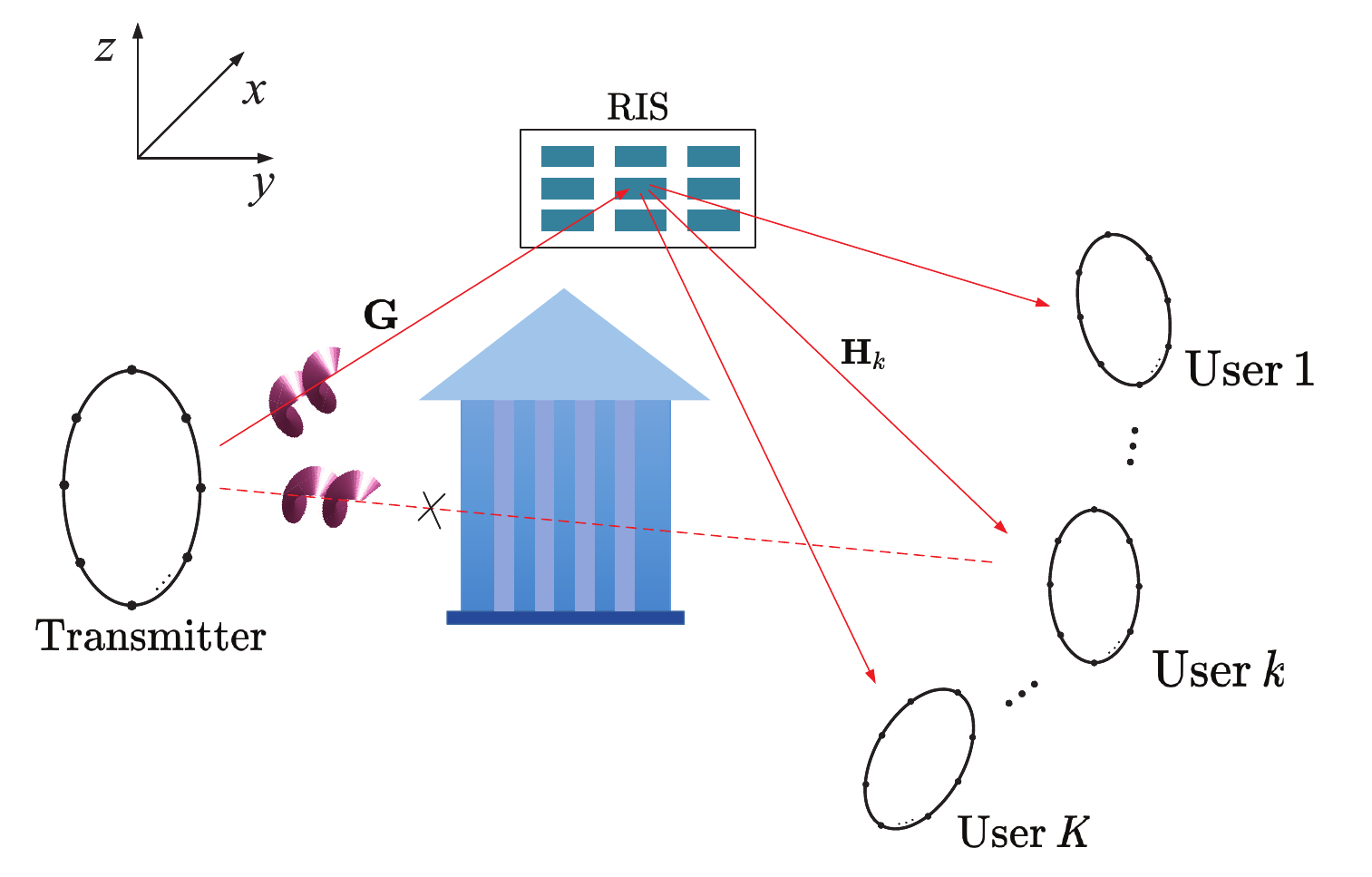}
\vspace{-0.4cm}
\caption{The MU-RIS-OAM system model.}
\label{fig:sys}
\end{figure}

We build the RIS-OAM-MU system as shown in Fig.~\ref{fig:sys}, where RIS is deployed to establish the direct links blocked by obstacles between the OAM transmitter and users. 
The OAM transmitter generated multiple OAM modes with $N_{\rm t}$-array UCA antenna sends signals to $K$ independent users. 
Thus, the total available OAM modes is equal to $N_{\rm t}$ for our developed MU-RIS-OAM system. 
To recover the received OAM signals, the $N_{\rm r}$-array UCA is utilized for each user. 
Assuming that $N_{k}$ OAM-modes are assigned to the $k$-th ($k = 1,2,\cdots, K$) user, we have $\sum_{k=1}^{K}N_{k} =N_{\rm t}$. Considering most practical communication scenarios, the UCA of OAM transmitter is misaligned to the UCAs of users. 
\vspace{-0.4cm}
\begin{figure}[hbtp]
\vspace{-0.2cm}
\centering
\includegraphics[width=0.44\textwidth]{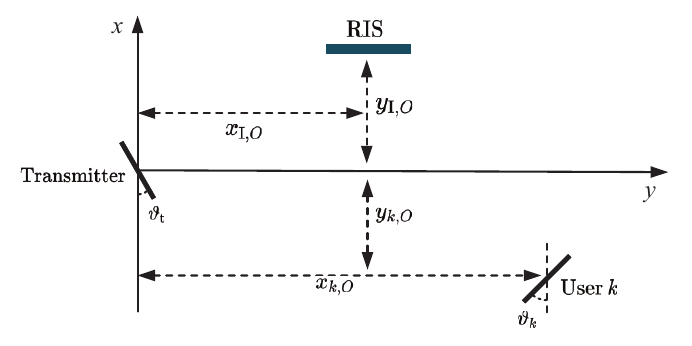}
\vspace{-0.3cm}
\caption{Positions of OAM transmitter, user $k$, and RIS in the xoy plane.}
\label{fig_coordinate}
\end{figure}
To obtain the channel gains, we first calculate the geometric positions of OAM transmitter, the $k$-th user, and RIS as shown in Fig.~\ref{fig_coordinate}, where RIS is assumed to be deployed on the yoz plane in the Cartesian coordinate. 
We denote by $\alpha_{{\rm t},n} =\alpha_{\rm t}(n-1)$ the angular position of the $n$-th $(n = 1,2,\cdots,N_{\rm t})$ antenna element on the OAM transmitter, where $\alpha_{\rm t} =2\pi/N_{\rm t}$ represents the phase difference between adjacent two antenna elements on the transmit UCA. 
Also, we denote by the $\bm{\omega}_{{\rm t},O}=[0,0,0]^T$, $\bm\omega_{k,O}=[x_{k,O},y_{k,O},0]^T$, and $\bm\omega_{{\rm RIS},O}=[x_{\rm I,\mathit O},y_{\rm I,\mathit O},0]^T$ the coordinates corresponding to the center points of the transmit UCA, the receive UCA of the $k$-th user, and RIS, respectively, where $(\cdot)^{T}$ represents the transpose of a matrix. 
Thus, we have the coordinates of all the antenna elements on the transmit UCA as $\bm\Omega_{\rm t} = [\bm\omega_{t,1},\cdots,\bm\omega_{t,N_t}]$, where the coordinate of the $n$-th antenna element on the OAM transmitter is expressed by
\begin{equation}
\label{e1}
\bm\omega_{\rm{t},\mathit{n}}=\bm{Q}_z(\vartheta_{\rm t}) R_{\rm t}[\cos(\alpha _{\rm{t},\mathit{n}}),0,\sin(\alpha _{\rm{t},\mathit{n}})]^T,
\end{equation}
where $R_{\rm t}$ represents the radius of the transmit UCA, $\vartheta _t =\arctan \big(\frac{x_{\rm I,\mathit O}}{y_{\rm I,\mathit O} }\big)$ denotes the angle of UCA's rotation around the z-axis, 
$x_{\rm I,O}$ and $y_{\rm I,O}$ represent denote the distance from the center points of RIS to the y-axis and x-axis respectively, 
$\bm{Q}_z(\cdot)$ is the rotation matrix  around the z-axis and $\bm{Q}_z(\vartheta_{\rm{t}})$ can be given by
\begin {equation}
\label{e2}
\bm{Q}_z(\vartheta_{\rm{t}}) = 
\begin{bmatrix}
 \cos{\vartheta_{\rm{t}} }  & \sin{\vartheta_{\rm{t}}} & 0\\
  -\sin{\vartheta_{\rm{t}}} & \cos{\vartheta_{\rm{t}}} & 0\\
 0 & 0 & 1
\end{bmatrix}.
\end{equation}
%


Likewise, we can get the coordinates of all the antenna elements on the receive UCA of the $k$-th user as $\bm\Omega_{k} = [\bm\omega_{k,1},\cdots,\bm\omega_{k,N_k}]$ and the coordinate of the $q$-th ($q=1,2,\cdots,N_{\rm r}$) antenna element on $k$-th user is expressed by
\begin{equation}
\label{e3}
\begin{split}
\bm\omega_{k,q} = \;\bm{Q}_z(-\vartheta_k) R_{k}[\cos(\alpha _{k,q}),0,\sin(\alpha _{k,q})]^T 
+ \bm\omega_{k,O},
\end{split}
\end{equation}
where $R_{k}$ is the UCA's radius of the $k$-th user, $\vartheta _k =\arctan \big(\frac{ x_{\rm I,\mathit O}-x_{k,O}}{y_{k,O}-y_{\rm I,\mathit O}}\big)$, $x_{k,O}$ and $y_{k,O}$ represent denote the distance from the center points of this UCA to the y-axis and x-axis respectively.

The RIS can be considered as a uniform rectangular array (URA) with $M=M_yM_z$ passive reflecting elements, $M_y$ and $M_z$  are the number of  elements on the RIS along the y-axis and z-axis, respectively.
Reflecting elements on the RIS can be expressed as follow: 
\begin{align}
\label{e4}
\bm\Omega_{\rm I}=\bm\omega_{{\rm RIS},O}\otimes\bm1_M^T + &\bigg[\bm0_M,d_y\left(\bm m_y \otimes \bm1_{M_z}+\dfrac{1-M_y}{2}\right),\nonumber\\
&d_z\left(\bm1_{M_z}\otimes\bm m_z  +\dfrac{1-M_z}{2} \right) \bigg]^T,
\end{align}
where $\bm m_y = [0,1,\cdots,M_y-1]^T$, $\bm m_z =  [0,1,\cdots,M_z-1]^T$, $\otimes$ denotes tensor product, 
$\bm0_M$ represents a $M$-by-$1$ vector of zero, 
$\bm1_{M_z}$ represents a $M_z$-by-$1$ vector of one,
$d_y$ and $d_z$  are the element separation distances on the RIS along the y-axis and z-axis, respectively.

The direct links  from the $n$-th antenna of the OAM transmitter to the $m$-th ($m = 1,\cdots, M$) passive reflecting element of RIS and from the $m$-th passive reflecting element of RIS to the $q$-th antenna of the receive UCA of the $k$-th user, denoted by $g_{mn}$ and $h_{qm,k}$, respectively, are expressed by
%
\begin{equation}
\label{e5_6}
\left\{\begin{matrix} 
g_{mn}=\dfrac{\beta\lambda}{4\pi {\left \| \bm\omega_{\rm I,\mathit m}- \bm\omega_{\rm{t},\mathit n} \right \|}} e^{-j{\tfrac{2\pi}{\lambda}}{\left \| \bm\omega_{\rm I,\mathit m}- \bm\omega_{\rm{t},\mathit n} \right \|} } \\
h_{qm,k}=\dfrac{\beta\lambda}{4\pi {\left \| \bm\omega_{k,\mathit q}- \bm\omega_{\rm I,\mathit m} \right \|}} e^{-j{\tfrac{2\pi}{\lambda}}{\left \|\bm\omega_{k,\mathit q} -\bm\omega_{\rm I,\mathit m}\right \|} }
\end{matrix}\right.,
\end{equation}
where $\lambda$ is the wavelength of the carrier signal, 
$\beta$ represents the attenuation, 
$\left \| \cdot \right \| $ denotes the Euclidean norm of a matrix, 
and $\bm\omega_{\rm I,\mathit m}$ is the $m$-th column of $\bm\Omega_{\rm I}$.

The transmit signal, denoted by $\mathbf x$, by the OAM transmitter is expressed as follows:
\begin{equation}
\label{e7}
\mathbf{x} = \frac{1}{\sqrt{N_{\rm t}}} \mathbf{W}_{\rm t} \mathbf{p}^{\frac{1}{2}} \mathbf s,
\end{equation}
where $\mathbf{W} _{\rm t}/{\sqrt{N_{\rm t}} }$ represents the inverse discrete Fourier transform (IDFT) matrix with $\textbf W_{\rm t}=[\mathbf w_{{\rm t},0},\cdots,\mathbf w_{{\rm t},N_{\rm t}}] \in \mathbb{C}^{N_{\rm t}\times N_{\rm t}}$, 
$\mathbf w_{{\rm t},l}=[e^{jl\alpha_{{\rm t},1}},\cdots, e^{jl\alpha_{{\rm t},N_{\rm t}}}]^T \in \mathbb{C}^{N_{\rm t}\times1}$ for the $l$-th ($0 \leq l \leq N_{\rm t}$) OAM mode, 
and $\mathbf{p}=\text{diag}[p_0,\cdots,p_l,\cdots,p_{N_{\rm t}-1}]$ denotes the power allocation diagonal matrix with respect to $p_{l}$ for the $l$-th OAM mode. 
Also, $\mathbf s \in \mathbb{C}^{N_{\rm t}\times 1}$ denotes the transmit modulated signals with the power constraint $\mathbb{E}(\bm s \bm s^H) = \mathbf I_{N_{\rm t} \times N_{\rm t}}$, where $\mathbb{E}(\cdot)$ represents the expectation operation and $\mathbf {I}_{N_{\rm t} \times N_{\rm t}}$ denotes the identity matrix with the dimension of $N_{\rm t} \times N_{\rm t}$.

After reflected by RIS, the received signal at the $k$-th user can be expressed as follows:
\begin {equation}
\label{e8}
 \mathbf y_k =  \mathbf H_{k} \boldsymbol{\Theta} \mathbf G \mathbf x + \mathbf n ,
\end{equation}
where $\mathbf H_{k}\in \mathbb{C}^{N_{\rm r} \times M}$ is the channel matrix with entries of $h_{qm,k}$ for the $k$-th user, 
$\mathbf G \in \mathbb{C}^{M\times N_{\rm{t}}}$ is the channel matrix with entries of $g_{mn}$, and  
$\mathbf n$ denotes the additive white Gaussian noise (AWGN) with variance $\sigma^2_k$ at the $k$-th user. 
Also, $\bm\Theta = \text {diag}(\bm\theta^H)$ represents the diagonal phase shift matrix of RIS with 
$\bm\theta = [\theta_1,\cdots,\theta_{m},\cdots,\theta_M]^T$, where $\theta_m = e^{j \varphi  _m}$  with $\varphi  _m \in [0,2\pi)$ denotes the phase shift of the $m$-th reflection element.

To decompose OAM signals, the discrete Fourier transform (DFT) is required. Hence, the decomposed signals, denoted by $\hat {\mathbf  y}_k$, can be derived as follows:
\begin {equation}
\label{e9}
 {\hat {\mathbf  y}}_k = \dfrac{1}{\sqrt{N_{\rm r} N_{\rm t}}} \mathbf W_k^H \mathbf H_{k} \bm\Theta \mathbf G   \mathbf{W}_{\rm t} \mathbf{p}^{\frac{1}{2}} \mathbf {s}     +{\tilde {\mathbf{n}}} ,
\end{equation}
where $(\cdot)^{H}$ represents the conjugate transpose of a matrix, 
$\mathbf W_k=[\mathbf w_{k,0},\cdots,\mathbf w_{k,N_{\rm r}}]$ denotes the DFT matrix with the dimension of $N_{\rm r} \times N_{\rm t}$,  
$\mathbf w_{{ k},l}=[e^{jl\alpha_{{k},1}},\cdots, e^{jl\alpha_{{k},N_{\rm r}}}]^T \in \mathbb{C}^{N_{\rm r}\times1}$ for the $l$-th OAM mode, 
and ${\tilde {\mathbf{n}}} = \frac{1}{\sqrt{N_{\rm t}}} \textbf W^H_k  \mathbf{n}$.

Based on Eq.~(\ref{e9}), the received signal-to-interference-plus-noise ratio (SINR), denoted by $\gamma_{l_{k}}$, of the $l$-th OAM mode for the $k$-th user can be derived as follows:
\begin {equation}
\label{e10}
\gamma_{l_k}=\frac{p_{l_k}{\left | u_{l_k,l_k} \right |}^2}{\sum\limits_{l_{j} =0, l_{j} \neq l_{k},}^{N_{\rm t}-1} p_{l_{j}}{\left | u_{l_{k},l_j} \right |}^2   + \sigma_k^2},
\end{equation}
where $u_{l_{k},l_j}=\frac{1}{\sqrt{N_{\rm r} N_{\rm t}}} \textbf w_{k,l_{k}}^H  \mathbf H_{k} \bm\Theta \mathbf G  \mathbf w_{{\rm t},l_j}$. 

The  weighted sum rate of all $K$ users for the MU-RIS-OAM system can be calculated as follows:
\begin {equation}
\label{e2_12}
C(\mathbf p,\bm\theta) = \sum_{k=1}^{K}\sum_{l_k=1}^{N_{k}} \omega_k  \log_{2}{(1+\gamma_{l_k})},
\end{equation}
%
where $\omega_k$ represents the weight factor. 
\section{Achieving Maximum ASR}\label{sec:ASR}
To maximize the ASR of our proposed  RIS-OAM-MU system, we jointly optimize the transmit power  allocation $\bf{p}$ for each user and the phase shifts  $\bm{\theta}$ of RIS subject to the constraints of total transmit power and the modulus of phase shifts   \cite{wu2019beamforming}.
Therefore, we formulate the optimization problem as  follows:
\begin{subequations}
\label{e12}
\begin{align}
\textbf{P1}:\underset{\mathbf p > \bm 0,\bm\theta }{\max } &~~ C(\mathbf p,\bm\theta) \label{12a} \\	
\mathrm{s.t.} &~~\text{tr}(\mathbf{p}^T\mathbf{p})\le P_{\rm t} , \label{12b} \\
&~~\left  | \theta _m \right | \le  1,\label{12c}
\end{align}
\end{subequations}
where $\text{tr}(\cdot)$ is the trace of a matrix and  
$P_{\rm t}$ denotes the total transmit power. 

To solve Problem $\textbf{P1}$, an efficient algorithm is required, 
which is challenging to be solved due to the coupling of variables in the objective function and the non-convexity of the unit-modulus constraints in Eq.~(\ref{12c}). 
The alternating optimization approach algorithm is developed to tackle Problem $\textbf{P1}$.

%
%
\subsection{Optimizing $\mathbf p$  with given ${\bm \theta}$} \label{sub1}
For given $\bm\theta$, Problem $\textbf{P1}$ can be rewritten as follows:
\begin{subequations}
\label{e13}
\begin{align}
\textbf{P2}:\; \underset{\mathbf p>0}{\max }  \; &~ C(\mathbf p,\bm\theta) \label{13a} \\	
\text{s.t.} \; &~\text{tr}(\mathbf{p}^T\mathbf{p})\le P_{\rm t}. \label{13b} 
\end{align}
\end{subequations}

To solve Problem $\textbf{P2}$, we first  convert the equivalent function according to the following theorem \cite{shen2018fractional2}.
\begin{theorem}
\label{tm}
Problem $\textbf{P2}$ is 
equivalent
 to Problem $\textbf{P3}$ given by 
\end{theorem}
\begin{subequations}
\label{e14}
\begin{align}
\textbf{P3}:\underset{\mathbf p>0, \bm \nu_{k}  }{\max }  \; &~ \sum _{k=1}^K f_k(\mathbf p,\bm\nu_k) \label{14a} \\	
\text{s.t.} \; &~\text{tr}(\mathbf{p}^T\mathbf{p})\le P_{\rm t}, \label{14b} 
\end{align}
\end{subequations}
where $\bm\nu_k = [\nu_{1_k},\cdots,\nu_{l_k},\cdots,\nu_{N_k}]^T $ are introduced as auxiliary variables corresponding to the received SINR term $\gamma_{l_{k}}$ and $f_k(\mathbf p,\bm\nu_k)$ is given by
\begin{multline}
\label{e15} 
f_k(\mathbf p,\bm\nu_k) = \sum_{l_k=1}^{N_{k}}  w_k\log_2 \left(1+\nu_{l_k}\right)-\sum_{l_k=1}^{N_{k}} w_k\nu_{l_k}\\
+\sum_{l_k=1}^{N_{k}} \frac{w_k(1+\nu_{l_k}) p_{{l_k}} |u_{{l_k},{l_k}}|^2}{\sum _{l_j = 1}^{N_{k}} p_{l_j} |u_{{l_k},{l_j}}|^2 +\sigma_k ^2}.  
\end{multline}

\begin{proof}
By substituting each SINR term with a new variable $\bm \nu_k$ \cite{shen2018fractional2}, Problem $\textbf{P2}$ can be reformulated as 
\begin{subequations}
\label{e16}
\begin{align}
\textbf{P4}:\underset{\mathbf p>0, \bm \nu_{k}}{\max }  \; &~ \sum_{k=1}^{K}\sum_{l_k=1}^{N_{k}} \omega_k  \log_{2}{(1+\nu_{l_k})} \label{16a} \\	
\text{s.t.} \; &~\text{tr}(\mathbf{p}^T\mathbf{p})\le P_{\rm t} , \label{16b} \\
&~ \nu_{l_k} \le \gamma_{l_k}.  \label{16c} 
\end{align}
\end{subequations}

The optimization presented above can be conceptualized as a two-level optimization, one involving an outer optimization over $\mathbf p$ and an inner optimization over $\nu_{l_k}$ with a fixed $\mathbf p$. The inner optimization is formulated as
\begin{subequations}
\label{e17}
\begin{align}
\textbf{P5}:\underset{\bm \nu_{k} }{\max }  \; &~ \sum_{k=1}^{K}\sum_{l_k=1}^{N_{k}} \omega_k  \log_{2}{(1+\nu_{l_k})} \label{17a} \\	
\text{s.t.} \; 
&~ \nu_{l_k} \le \gamma_{l_{k}}.  \label{17b} 
\end{align}
\end{subequations}
The solution to this inner optimization is obviously that $\nu_{l_k}$
should satisfy Eq.~(\ref{17b}) with equality.

Since Problem $\textbf{P5}$ is a convex optimization in $\bm\nu_k$, the strong duality holds \cite{shen2018fractional2}. 
We can introduce the dual variable $\bm\lambda_k = [\lambda_{1_k},\cdots,\lambda_{l_k},\cdots,\lambda_{N_k}]^T $ for each inequality constraint in Eq.~(\ref{17b}) and then form the Lagrangian function as follows:
\begin {multline}
\label{e18}
\mathcal{L}(\bm\nu_k, \bm\lambda_k) = \sum_{k=1}^{K}\sum_{l_k=1}^{N_{k}} w_k\log_2 \left(1+\nu_{l_k}\right) \\ - \sum_{k=1}^{K}\sum_{l_k=1}^{N_{k}} \lambda_{l_{k}}(\nu_{l_{k}} - \gamma_{l_{k}}).
\end{multline}

Due to strong duality, Problem $\textbf{P5}$ is equivalent to the dual problem  $\underset{\bm\lambda_{k} \succeq 0}{\min } \underset{\bm\nu_{k} }{\max }~\mathcal{L}(\bm\nu_{k}, \bm\lambda_{k})$.
By setting $\partial \mathcal{L} / \partial \nu _{l_{k}} = 0$, 
we have the optimal solution of $\lambda_{l_{k}}$, denoted by $\lambda_{l_{k}} ^\star = w_k /{1 + \nu_{l_{k}} ^\star}$.
Thus,  $\lambda_{l_{k}} ^\star$ can be derived as follows:
\begin {equation}
\label{e20}
\lambda_{l_{k}} ^\star = w_k \frac{\sum_{k=1}^{K}
\sum_{l_{j} =1, l_{j} \neq l_{k}}^{N_{k}} p_{l_{j}}{\left | u_{l_{k},l_j} \right |}^2   + \sigma_k^2
}{\sum_{k=1}^{K}\sum _{l_j = 1}^{N_{k}} p_{l_j} |u_{{l_k},{l_j}}|^2 +\sigma_k ^2}.
\end{equation}
It can be easily found that Eq.~(\ref{e15}) becomes a differentiable concave function over $\bm\nu_k$ with given $\mathbf p$. Thus, the optimal solution of $\nu_{l_{k}}$, denoted by $\nu_{l_{k}}^{\star}$, can be obtained by setting $\partial f_k / \partial \nu_{l_{k}} = 0$.
Then, substituting $\nu_{l_{k}}^\star$ into $f_k(\mathbf p,\bm\nu_k)$, the objective function of Eq.~(\ref{13a}) is equal to $f_k(\mathbf p,\bm\nu_k)$. 
Thus, the equivalence of Problems $\textbf{P2}$ and $\textbf{P3}$ are established.
\end{proof}
According to \emph{Theorem~1} and given $\mathbf p$ in Eq.~(\ref{e15}), we can obtain $\nu_{l_{k}}^{\star}$ by setting $\partial f_k / \partial \nu_{l_k}$ to zero,
then fixing $\nu_{l_k}$ and dropping irrelevant constant terms, 
$f_k(\mathbf p,\bm\nu_k)$ can be expressed as follows: 
\begin {equation}
\label{e24}
f_k'(\mathbf p,\bm\nu_k) = \sum_{k=1}^{K}\sum_{l_k=1}^{N_{k}} \frac{w_k(1+\nu _{l_k}) p_{l_k} |u_{{l_k},{l_k}}|^2}{\sum_{k=1}^{K}\sum _{l_j = 1}^{N_{k}} p_{l_j} |u_{{l_k},l_j}|^2 +\sigma_k ^2},
\end{equation}
which is a sum-of-ratio form.  
Aiming at obtaining the optimal solutions of $\mathbf p$,  we use the quadratic transformation (QT) method to rewrite  Eq.~(\ref{e24}) as follows:
\begin{multline} 
\label{e25}
\tilde f_k(\mathbf {p},\boldsymbol \nu_k,\boldsymbol \eta_k) = \sum_{k=1}^{K}\sum _{l_k = 1}^{N_{k}} 2\eta_{l_k}\sqrt{w_k(1+\nu _{l_k})p_{{l_k}} |u_{{l_k},{l_k}}|^2}\\ 
- \sum_{k=1}^{K}\sum _{l_k = 1}^{N_{k}} \eta^2_{l_k} \Bigg (\sum_{k=1}^{K}\sum _{l_j = 1}^{N_{k}}p_{l_j} |u_{{l_k},{l_j}}|^2 + \sigma_k ^2\Bigg) ,
\end{multline}
and $\boldsymbol \eta_k = [\eta_{1_k},\cdots,\eta_{l_k},\cdots,\eta_{N_k}]^T $ is the auxiliary variables.  

Then, we have the optimal solutions, denoted by $\eta_{l_{k}}^{\star}$ and $p_{l_k}^{\star}$, of $\boldsymbol \eta_k$ and $\mathbf {p}$ by $\partial \tilde f_k / \partial \eta_{l_k} = 0$ and $\partial \tilde f_k / \partial p_{l_k} = 0$, respectively, with fixing other variables\cite{shen2018fractional2}. Hence, we have the closed-form expressions of $\eta_{l_{k}}^{\star}$ and $p_{l_k}^{\star}$ given by
\begin{equation}
\label{e26_27}
\left\{\begin{matrix} 
\eta^\star_{l_k} = \frac{\sqrt{w_k( 1 + \nu _{l_k}) p_{l_k} |u_{{l_k},{l_k}}|^2}}{\sum_{k=1}^{K}\sum _{l_j = 1}^{N_{k}} p_{l_j} |u_{{l_k},{l_j}}|^2+\sigma _k^2} \\ p^\star _{l_k} = \min \Bigg \lbrace P_{\rm t},~ \frac{\eta^2_{l_k} w_k (1+\nu _{l_k})|u_{{l_k},{l_k}}|^2}{\big (\sum_{k=1}^{K}\sum _{l_j = 1}^{N_{k}} \eta^2_{l_j} |u_{{l_j},{l_k}}|^2\big)^2}\Bigg  \rbrace
\end{matrix}\right. .
\end{equation}
\subsection{Optimizing ${\bm \theta}$  with given $\mathbf p$} \label{sub2}
With given $\mathbf p$, Problem $\textbf{P1}$ can be formulated as follows:
\begin{subequations}
\label{e28}
\begin{align}
\textbf{P6}:\underset{\bm\theta }{\max }  \; &~ C(\mathbf p,\bm\theta) \label{28a} \\	
\text{s.t.} \; &~\left  | \theta _m \right | \le  1. \label{28b} 
\end{align}
\end{subequations}

By applying the change of variables $u_{{l_k},{l_j}}=\bm\theta^H \textbf a_{{l_k},{l_j}}$ where  $\textbf a_{{l_k},{l_j}} = \frac{1}{\sqrt{N_{\rm r} N_{\rm t}}} \text{diag}(\mathbf w_{k,l_{k}}^H  \mathbf H_{k}) \mathbf G  \mathbf w_{{\rm t},l_j}$,
and according to \emph{Theorem~1}, 
we introduce auxiliary variables $\boldsymbol {\tilde \nu}_k$. 
Thereby, Eq.~(\ref{28a}) can be converted to $g_k'(\bm\theta ,\boldsymbol {\tilde \nu}_k)$ given as follows:
\begin {equation}
\label{e32}
g_k'(\bm\theta,\bm{\tilde\nu}_k) = \sum_{k=1}^{K}\sum _{l_k = 1}^{N_{k}} \frac{w_k(1+\tilde\nu _{l_k}) p_{{l_k}} |\bm\theta^H \textbf a_{{l_k},{l_k}}|^2}
{\sum_{k=1}^{K}\sum _{l_j = 1}^{N_{k}} p_{l_j} |\bm\theta^H \textbf a_{{l_k},{l_j}}|^2 +\sigma_k ^2}.
\end{equation}
%
We then transform Eq.~(\ref{e32}) by using the QT method to
\begin{multline} 
\label{e33}
\tilde g_k(\bm\theta,\boldsymbol {\tilde\nu}_k,\boldsymbol{ \tilde \eta}_k) = \sum_{k=1}^{K}\sum _{l_k = 1}^{N_{k}} 2\eta_{l_k}\sqrt{w_k(1+\tilde\nu _{l_k})p_{{l_k}} |\bm\theta^H \textbf a_{{l_k},{l_k}}|^2}\\ 
- \sum_{k=1}^{K}\sum _{l_k = 1}^{N_{k}} \eta^2_{l_k} \Bigg ( \sum_{k=1}^{K}\sum _{l_j = 1}^{N_{k}} p_{l_j} |\bm\theta^H \textbf a_{{l_k},{l_j}}|^2 + \sigma_k ^2\Bigg) ,
\end{multline}
where $\boldsymbol { \tilde \eta}_k = [{ \tilde \eta}_{1_k},\cdots,{ \tilde \eta}_{l_k},\cdots,{ \tilde \eta}_{N_k}]^T $ is the auxiliary variables.
Dropping all the irrelevant constant terms of $\tilde g_k(\bm\theta,\boldsymbol {\tilde\nu}_k,\boldsymbol{ \tilde \eta}_k)$, Problem $\textbf{P6}$ can be further expressed by 
\begin{subequations}
\label{e35}
\begin{align}
\textbf{P7}:\underset{\bm\theta}{\min }  \; &~ \hat g_k(\bm\theta,\boldsymbol {\tilde\nu}_k,\boldsymbol{ \tilde \eta}_k) \label{35a} \\	
\text{s.t.} \; &  \left  | \theta _m \right | \le  1,\label{35b} \\
 \hat g_k (\bm\theta,\boldsymbol {\tilde\nu}_k,\boldsymbol{ \tilde \eta}_k) &= \bm\theta^H \textbf U \bm\theta - 2 \mathfrak{R}\{ \bm\theta^H \textbf v \},  \\
\textbf U = \sum_{k=1}^{K}\sum _{l_k = 1}^{N_{k}} &{ \tilde \eta}^2_{l_k} \Bigg ( \sum_{k=1}^{K}\sum _{l_j = 1}^{N_{k}} p_{{l_j}} \textbf a_{{l_k},{l_j}}\textbf a_{{l_k},{l_j}}^H  \Bigg ),  \\
\textbf v =\sum_{k=1}^{K}\sum _{l_k = 1}^{N_{k}} &2{ \tilde \eta}_{l_k}\sqrt{w_k(1+{\tilde\nu} _{l_k})p_{{l_k}}  }   \textbf a_{{l_k},{l_k}},
\end{align}
\end{subequations}
%
and  $\mathfrak{R} \{ \cdot \}$ denotes the real part of a complex number.

\begin{algorithm}[htbp]
\caption{}
\label{alg1} 
\begin{algorithmic}
\STATE 1:{$\mathbf{ Initialize:}$ $\mathbf p^{(0)}$, $\bm\theta^{(0)}$}, and $t=0$
\STATE 2:{$\mathbf{Repeat :}$ 

\qquad $\mathbf{ Step 1:}$ Update $\boldsymbol \nu ^{(t)}_{k}$, $\boldsymbol {\tilde\nu} ^{(t)}_{k}$, $\boldsymbol{ \eta}^{(t)}_{k}$ and $\boldsymbol{ \tilde \eta}^{(t)}_{k}$  with given $\mathbf p^{(t)}$ and $\bm\theta^{(t)}$;



\qquad $\mathbf{ Step 2:}$ Update $\mathbf p^{(t+1)}$ by Eq.~(\ref{e26_27}) with given $\boldsymbol \nu ^{(t)}_{k}$, $\boldsymbol{ \eta}^{(t)}_{k}$, $\mathbf p^{(t)}$, and $\bm\theta^{(t)}$;

\qquad $\mathbf{ Step 3:}$  Update $\bm\theta ^{(t+1)}$ by Eq.~(\ref{e35}) with given $\boldsymbol {\tilde\nu} ^{(t)}_{k}$, $\boldsymbol{ \tilde \eta}^{(t)}_{k}$, $\mathbf p^{(t+1)}$, and $\bm\theta^{(t)}$;

\qquad $\mathbf{ Step 4:}$ $t=t+1$;
}

\STATE 3:{$\mathbf{ Until :}$ the value of Problem $\textbf{P1}$ converges. }
\end{algorithmic}
\end{algorithm}
It is obvious that Problem $\textbf{P7}$ is a convex optimization problem, which can be solved by convex optimization solvers such as CVX. $\textbf{Algorithm~\ref{alg1}}$ summarizes the entire alternating algorithm of Problem $\textbf{P1}$.
\section{Numerical Results}\label{sec:sim}
Throughout the whole simulation, we set $K=3$, $N_{\rm r}$ = 5, $N_{\rm t}$ = 15, $R_{\rm t}$ = $R_{k}$ = 0.6 m, $\alpha_{\rm t,0} = \alpha_{k,0} = 0^{\circ}$, and the carrier frequency as 10 GHz. Also, we set $x_{\rm I,O} =$ 2 m, $x_{k,O} =$ 0 m, $y_{\rm I,O} = $ 30 m, $y_{k,O} =$ 20 m, and the distance between users as 10 m. The number of passive reflecting elements RIS is $M  = 40$ and  $P_{\rm t}$ is set as 20 dB.

\begin{figure}[htbp]
\centering
\includegraphics[width=0.45\textwidth]{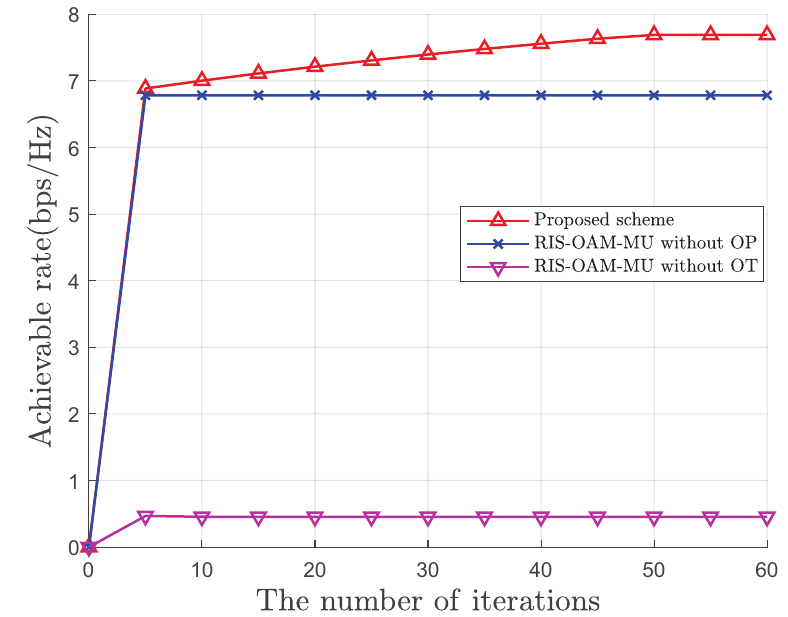}
\caption{Achievable rate versus the number of iterations.}
\label {fig3}
\end{figure}

\begin{figure}[htbp]
\centering
\includegraphics[width=0.45\textwidth]{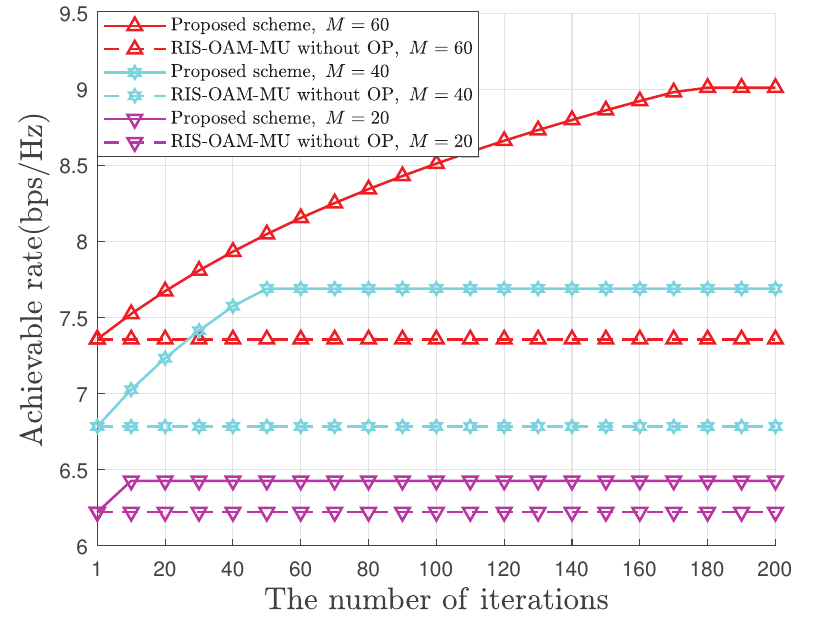}
\caption{Achievable rate versus different $M$ of RIS.}
\label {fig5}
\end{figure}

Figure~\ref{fig3} compares the convergence of our proposed scheme and several other schemes, where RIS-OAM-MU without OP scheme only optimize the transmit power allocation for the RIS-OAM-MU scheme, RIS-OAM-MU without OT scheme only optimizes the phase shifts of RIS.  It is observed that our proposed scheme is significantly superior to other schemes and converges after approximately 50 iterations. This is because OAM has high degrees of freedom and the inter-user interference can be suppressed with our proposed scheme.

To verify the impact of the number of reflecting elements of RIS, Fig.~\ref{fig5} presents the ASR with different $M$, where we set $M=60,40$, and 20. It is clear that the ASR increases as $M$ increases. The reason is that the strength of  received signals is enhanced.
\begin{figure}[htbp]
\centering
\includegraphics[width=0.48\textwidth]{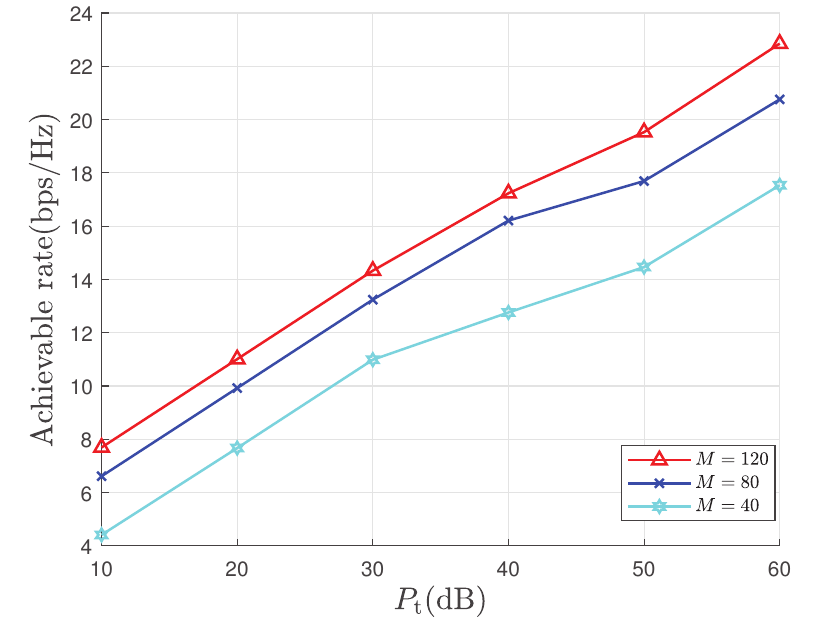}
\caption{Achievable rate versus $P_{\rm t}$.}
\label {fig4}
\end{figure}

Figure~\ref{fig4} presents the ASR of our proposed scheme versus the total transmit power $P_{\rm t}$, where we set $M = $ 120, 80, and 40.
It is shown that designing the RIS’s phase shifts, the strength of received  signals  can be enhanced, thus significantly increasing the ASR of MU wireless communications. Also, it can be observed that the ASR increases as $P_{\rm t}$ increases. This is due to the enhancement of signal.
\section{Conclusion} \label{sec:conc}
In this paper, 
the RIS-OAM-MU scheme was proposed to significantly increase the ASR of wireless communication scenarios where the direct links between the OAM transmitter and multiple users are blocked by surroundings. 
The alternative optimization algorithm was developed to joint optimize the transmit power allocation to each OAM mode and the phase shifts of RIS, thus resulting in maximizing the ASR with low computational complexity. 
The numerical outcomes have demonstrated that the employment of  RIS in MU-OAM wireless communications along with our developed optimization algorithm can enhance the ASR in the presence of direct link blockages.


\vspace{12pt}


\begin{thebibliography}{00}
\bibitem{9003618}
W. Jiang, B. Han, M. A. Habibi and H. D. Schotten, ``The Road Towards 6G: A Comprehensive Survey,'' \emph{IEEE Open Journal of the Communications Society}, vol. 2, pp. 334--366, 2021.
  
  \bibitem{1} W. Cheng, Y. Xiao, S. Zhang and J. Wang, “Adaptive Finite Blocklength for Ultra-Low Latency in Wireless Communications,” \emph{IEEE Transactions on Wireless Communications}, vol. 21, no. 6, pp. 4450--4463, June 2022.

\bibitem{2} H. Jing, W. Cheng and X. -G. Xia, “Fast Transceiver Design for RIS-Assisted MIMO mmWave Wireless Communications,” \emph{IEEE Transactions on Wireless Communications}, vol. 22, no. 12, pp. 9939--9954, Dec. 2023.

\bibitem{liang2020joint} L.~Liang, W.~Cheng, W.~Zhang, and H.~Zhang, ``Joint OAM Multiplexing and OFDM in Sparse Multipath Environments,'' \emph{IEEE Transactions on Vehicular  Technology}, vol. 69, no. 4, pp. 3864--3878, April 2020.
  
  \bibitem{3}
  R. Lyu, W. Cheng, M. Wang, and W. Zhang, ``Fractal OAM Generation and Detection Schemes,'' \emph{IEEE Journal on Selected Areas in Communications}, vol. 42, no. 6, pp. 1598--1612, June 2024.

\bibitem{9442905}
W. -X. Long, R.~Chen, M.~Moretti, J.~Xiong, and J.~Li, ``Joint Spatial Division and Coaxial Multiplexing for Downlink Multi-User OAM Wireless Backhaul,''
  \emph{IEEE Transactions on Broadcasting}, vol. 67, no. 4, pp. 879--893, Dec. 2021.

\bibitem{yang2021reconfigurable} Z.~Yang, Y.~Hu, Z.~Zhang, W.~Xu, C.~Zhong, and K.-K. Wong, ``Reconfigurable Intelligent Surface Based Orbital Angular Momentum: Architecture, Opportunities, and Challenges,'' \emph{IEEE Wireless Communications},
  vol. 28, no. 6, pp. 132--137, Dec. 2021.


\bibitem{10330152}
Y.~Chen, W.~Cheng, and W.~Zhang, ``Reconfigurable Intelligent Surface Equipped UAV in Emergency Wireless Communications: A New Fading–Shadowing Model and Performance Analysis,'' \emph{IEEE Transactions on Communications}, vol. 72, no. 3, pp. 1821--1834, March 2024.

\bibitem{wu2019beamforming} Q.~Wu and R.~Zhang, ``Beamforming Optimization for Wireless Network Aided by Intelligent Reflecting Surface With Discrete Phase Shifts,'' \emph{IEEE
  Transactions on Communications}, vol. 68, no. 3, pp. 1838--1851, March 2020.

\bibitem{shen2018fractional2}
K.~Shen and W.~Yu,  ``Fractional Programming for Communication Systems—Part II: Uplink Scheduling via Matching,'' \emph{IEEE Transactions on Signal Processing},
  vol. 66, no. 10, pp. 2631--2644, May 2018.


\end{thebibliography}
\end{document}